\documentclass[twoside]{mhd}

\usepackage{epsfig}
\usepackage{graphicx}
\title{Parker's model in geodynamo}

\author{M. Reshetnyak}

 \institute{Institute of the  Physics of the Earth, B.Gruzinskaya, 10,
Moscow, 123995, Russia, m.reshetnyak@gmail.com
} 

\begin{document}
\maketitle


\begin{abstract} 
    We consider how information on geostrophic flows in the planetary cores, taken from 3D simulations in the sphere,  can be used in 2D Parker's geodynamo model with the simple forms of the $\alpha$-quenching. 
     Using cluster computer systems  dependence of  dynamo equations solution  on the magnitudes of $\alpha$- and $\omega$-effects is studied. We show that geostrophical flow can produce the well-known Z-structure of the
      poloidal magnetic field without the feed-back of the magnetic field onto the large-scale velocity field. 
      The influence of fluctuations of $\alpha$-effect on magnetic field generation, its spectral properties, in respect to the  geodynamo applications,  is also discussed.
 \end{abstract}


\section*{Introduction}

Parker's dynamo \cite{Parker:ApJ:122:293} equations, which can be rigorously derived \cite{Fioc:Magn:31:18} from the general mean-field dynamo equations
\cite{KR:book} is a good candidate for the various physical applications, starting from the  galactic dynamo to the dynamos in the
   planets. The details of the particular equations  depend on our knowledge on  these objects.

 Originally the
$\alpha$-effect and  the differential rotation, $\omega$,  were the prescribed functions of spatial coordinates,
  and could vary from model to  model.  It was clear that without  additional information on
the hydrodynamics of the system this approach   could  produce only some very   general features of the dynamo mechanism. The further
   specification  of Parker equations  is concerned with the derivation of the tensor  forms of  $\alpha$,  
 the differential rotation curve, and study of   the feedback mechanism 
  of the magnetic field onto the flow \cite{RKH:13}. As regards to the latter point, the first naive idea was that the magnetic field suppresses the turbulence, and as a result, 
 $\alpha$ is   proportional to the inversed magnetic energy $E_m$.   
 The quenching  can be local, when the magnetic energy is simulated in the particular physical point, either it is global, with  $E_m$ averaged 
   over the whole volume.  The both variants are named as the algebraic $\alpha$-quenching.

The more sophisticated, dynamic quenching \cite{KRR:95}, is based on the
idea that there is a magnetic contribution  $\alpha_m$
 to the total $\alpha$,   which is governed by the evolutionary differential equation.  This kind of the
  non-linearity  also leads to the saturation of the dynamo  system,         however the details of the process can differ from the
algebraic model.
 The both models of quenching are still used  in the galactic and solar  dynamos, where the magnetic energy of the mean field is compared on the order of
  magnitude with the kinetic energy of the turbulence, being  much less than the kinetic energy of the large-scale flow. 

  The other form  of the magnetic field quenching is  modification of the large-scale flow with the mean magnetic field. Thus, the
  estimates of the velocities in the Earth's core, based on the west-drift velocity of the magnetic field, resulted in a
    conclusion that the magnetic energy is the  three order of magnitude  larger than the kinetic energy in the system of coordinates of
 the rotating mantle of the Earth.  This  effect is caused by  the rapid rotation of the planet. It was the argument that the
  magnetic field should modify the large-scale flow as well, and as a result, stimulated development of the full dynamo system equations with the
   large-scale Lorentz force included in the Navier-Stokes equation. This scenario was realized in the remarkable Z-model of Braginsky \cite{Br:75},
 where equations for the axi-symmetrical flow were solved
  explicitly. In fact, equation of motion described the balance   of the Lorentz, Coriolis and Archemedean forces, and the 
   pressure gradient. Viscosity was neglected in the main volume and took into account only in the boundary layers. From the 
  numerical point of view Z-model was quite complicated, because due to geostrophy simulation  of the Navier-Stokes equation required
 integration on z-direction,
  and one should switch from the spherical coordinates to the cylindrical at the every time step.
  
     The other, may be  the most difficult  point in that time, was  absence of information on the hydrodynamic of the flow in
    the core. It was realized by the mean-field and geodynamo communities only latter   that convection has a cyclonic form  
  \cite{Roberts:1968,Busse:1970},  and that its
  parametrization in terms of the
    mean-field dynamo is a very tricky point. For example, for the Rayleigh numbers $\rm Ra$, close to the critical one,
  convection has cyclonic form without differential rotation.
        Then, in terms of the mean-field  dynamo it should be the $\alpha^2$-dynamo. Note that the Rossby number is very small, ${\rm Ro}\ll 1$, and the regime of $\alpha\omega$-dynamo was expected. The further increase of 
  $\rm Ra$ causes the Rossby waves \cite{Busse:2002}, the cyclones start to rotate around the axis of rotation, and as a result, the differential rotation
   appears. This case already corresponds to the 
   $\alpha^2\omega$- or  $\alpha\omega$-regimes. It would be very useful in this situation to use external information on the flow from 3D simulations.
   
  Having this in mind, we  consider the simplest case of the mean-field equations with the algebraic $\alpha$-quenching, and stationary  $\omega$.
   Here we show that using information on the geostrophic flows, derived from 3D simulation in the rotating convective shell, we can
 produce Z-structure of the poloidal magnetic field in 2D  Parker's model. We also discuss in what
    extent the random fluctuations of the $\alpha$-effect can be used for the modeling of the geomagnetic reversals. So as the result is very sensitive  to the magnitudes of the $\alpha$- and $\omega$-effects, $C_\alpha$, $C_\omega$, which are pure known, we show how the main parameters, such as the intensity of the dipole, magnitude of its oscillations, position of the maximum in the spectrum of the Legendre polynomials  depend on $C_\alpha$, and $C_\omega$.


\section{Parker's  equations and numerical methods}

\label{section:2}
The mean magnetic field $\bf B$  is governed by the induction equation
\begin{equation}\label{Parker:1}
{\partial{\bf
B}\over\partial t}=\nabla \times \Big( \alpha\,{\bf B}+
{\bf V}\times {\bf B}
-\eta\, {\rm rot}{\bf B}   \Big), 
\end{equation}
where $\bf V$ is the large-scale velocity field,  $\alpha$ is the $\alpha$-effect,  and $\eta$ is a magnetic diffusion.
 The  magnetic field ${\bf B}=\left( {\bf B^p},\, {\bf B^t} \right)$
has two parts: the poloidal component ${\bf B^p}=\nabla\times {\bf A}$, 
where $\bf A$ is the  vector  potential of the magnetic field, and the toroidal component $\bf B^t$.

 In the axi-symmetrical case  the vector potential $\bf A$ and $\bf B^t$ have 
the only one azimuthal component
 in the spherical system of coordinates
$(r,\,\theta,\, \varphi)$: ${\bf A}(r,\,\theta)=(0,\, 0,\, A)$, and ${\bf B^t}(r,\,\theta)=(0,\, 0,\, B)$. 

The poloidal field  can be written in the form: 
\begin{equation}\label{Parker:2}
\displaystyle
{\bf B^p}=
\left(
 {1\over r\, \sin\theta}{\partial\over\partial \theta }\left( A\, \sin\theta \right),\,
-{1\over r} {\partial \over \partial r}  \left( r\, A \right),\, 0
 \right).
\end{equation}

In terms of scalars $A$ and $B$ Eq(\ref{Parker:1}) is reduced to the following system of equations:
\begin{equation}\label{Parker:3}
\begin{array}{l}
\displaystyle
{\partial{
A}\over\partial t}=\alpha {B} + \left({\bf V}\times\,{\bf B}\right)_\varphi
+\eta \left( \nabla^2 - {1\over r^2\sin^2\theta}  \right) {A}
\\  \\ 
\displaystyle
{\partial{B}\over\partial t}={\rm rot}_\varphi \left( \alpha\,{\bf B} +
 {\bf V}\times\,{\bf B}\right)+\eta
 \left( \nabla^2 -{1\over r^2\sin^2\theta}  \right){B},
\end{array}
\end{equation}
where  the subscript $\varphi$ corresponds to the azimuthal component of the vector.

Eqs(\ref{Parker:3}),   solved in the spherical shell
$r_i\le r\le r_\circ$ with $r_i=0.35$, $r_\circ=1$ typical for the Earth's core,
 are closed with the pseudo-vacuum boundary conditions: ${ B}=0$, and $\displaystyle {\partial \over\partial r} \left( r A\right)=0$ at $r_i$ and $r_\circ$
 and $A=B=0$ at the axis of rotation $\theta=0,\, \pi$. The simplified form of the vacuum boundary condition  for $A$ is  well adopted in dynamo community, and presents a good approximation of the boundary with the non-conductive medium 
\cite{Jouve:2008}. The reason why
 the vacuum boundary condition is  used at the inner core boundary is  concerned with the weak influence of the inner core on the reversals statistics of the magnetic field \cite{Wicht:2002}.

  In the general case velocity  $\bf V$ is a three-dimensional vector, as a function of $r$ and $\theta$. Further we consider only the effect of the differential rotation, concerned with the $\varphi$-component of $\bf V$, leaving the input of the meridional circulation  $(V_r,\,V_\theta)$ out of the scope of the paper.
The amplitude of $V_\varphi$ is defined by constant $C_\omega$.

The model is closed with the  alpha-quenching in the local algebraic form:
  \begin{equation}\label{Parker:4}\displaystyle
     \alpha=C_\alpha\,{\alpha_\circ  \over 1+E_m(r,\,\theta)},
 \end{equation}
where $E_m$ is the magnetic energy, and $C_\alpha$ is a constant.

    The system (\ref{Parker:3},\ref{Parker:4}) was solved using the $4^{th}$ order Runge-Kutta method, where  the r.h.s. derivatives were   approximated 
 with the second-order central-differences. 
  These algorithms resulted in  C++ object oriented code with use of Blitz C++ library for the easier compact operations with the arrays. The  post-processor graphic visualization was organized using the Python graphic library MatPlotlib. All simulations were done under the Ubuntu  OS. 
 See the details of the benchmarks in \cite{Reshetnyak:2014}.

        To demonstrate dependence of solution of Eqs(\ref{Parker:3}, \ref{Parker:4}) on  parameters ($C_\alpha$,  $C_\omega$)  the  MPI wrapper was used   to run 
    the main program at two  cluster supercomputers:  Lomonosov in Moscow State University and at the Joint Supercomputer Center of RAS. 
     The wrapper called the main program  with the fixed different values of  
 ($C_\alpha$,  $C_\omega$) at the different processors and then gathered all the data after the end of simulations. Usually, 10 different values of  
$C_\alpha$ and  $C_\omega$ were used, and as a result, 100 processors were needed. The one additional processor was used for synchronization.

\section{Numerical results} 
Convection in the planetary cores is characterized by the very rapid rotation, so that the Rossby number is small, $\rm Ro\ll 1$. 
 As a result, the so-called geostrophic balance in the bulk of the liquid core takes place.  The geostrophic state corresponds to  the balance of the pressure gradient and Coriolis force \cite{Pedlosky:1987}. 
   In this state temperature fluctuations, velocity $\bf V$, and the kinetic helicity $\chi={\bf V}\cdot {\rm rot}{\bf V}$ 
 have large gradients in directions perpendicular to the axis of rotation $\bf z$, and weak dependence along $z$-coordinate. 
 Having in mind  that the magnetic energy  
      has small effect on the large-scale flow \cite{Jones:2000}, for the mean-field equations (\ref{Parker:3}) we can use averaged $\bf V$, and $\alpha$,
       taken from the non-magnetic 3D simulations of convection. For this aim simulations in the spherical shell for the Boussinesq convection with the  stress-free boundary condition for $\bf V$, and fixed temperatures at the boundaries  \cite{Reshetnyak:2010}, were used.

\begin{figure}[th!]
\def \ss {7cm}
\vskip -1cm
\includegraphics[width=\ss]{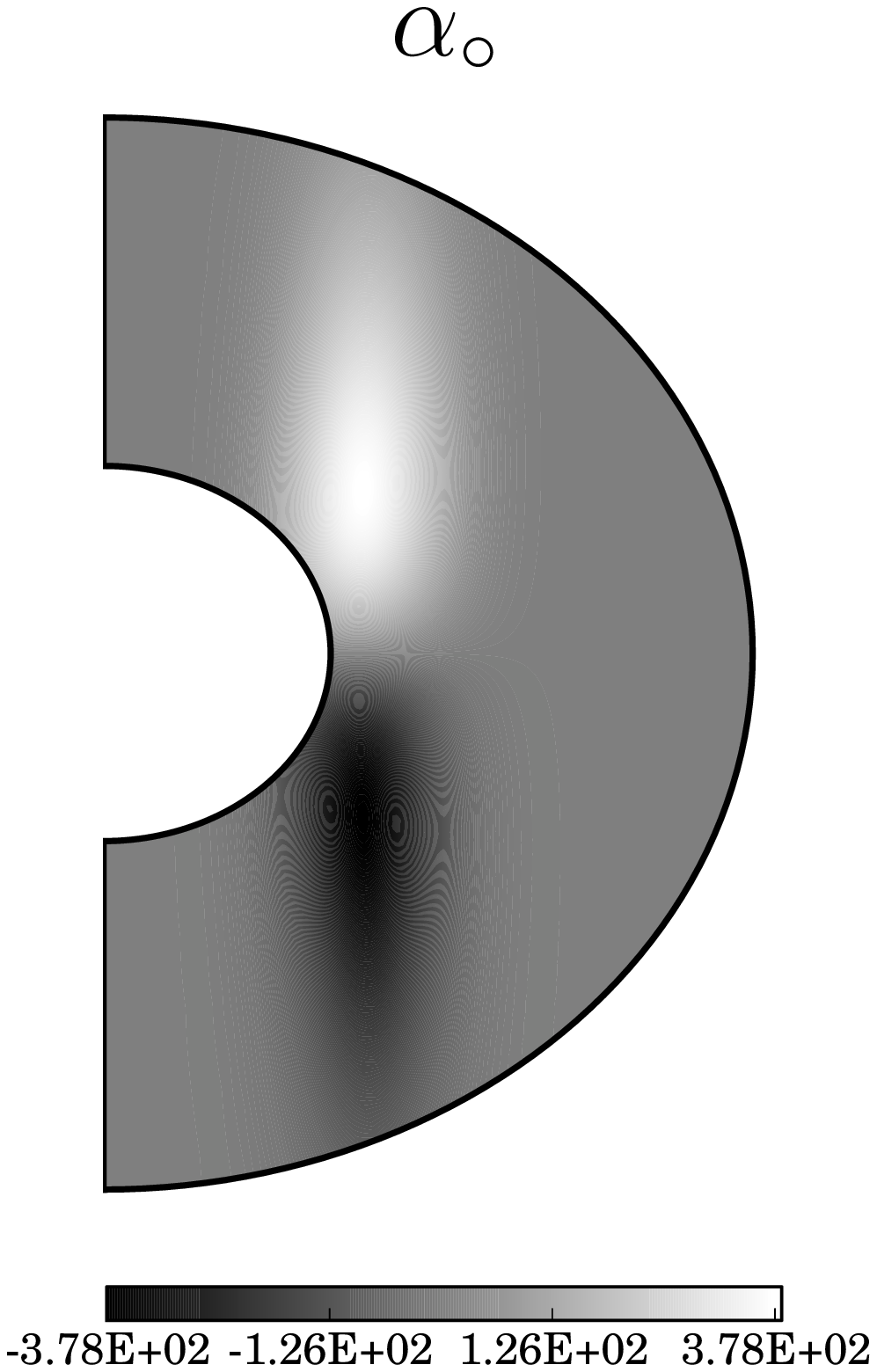}
\hskip -2cm 
\includegraphics[width=\ss]{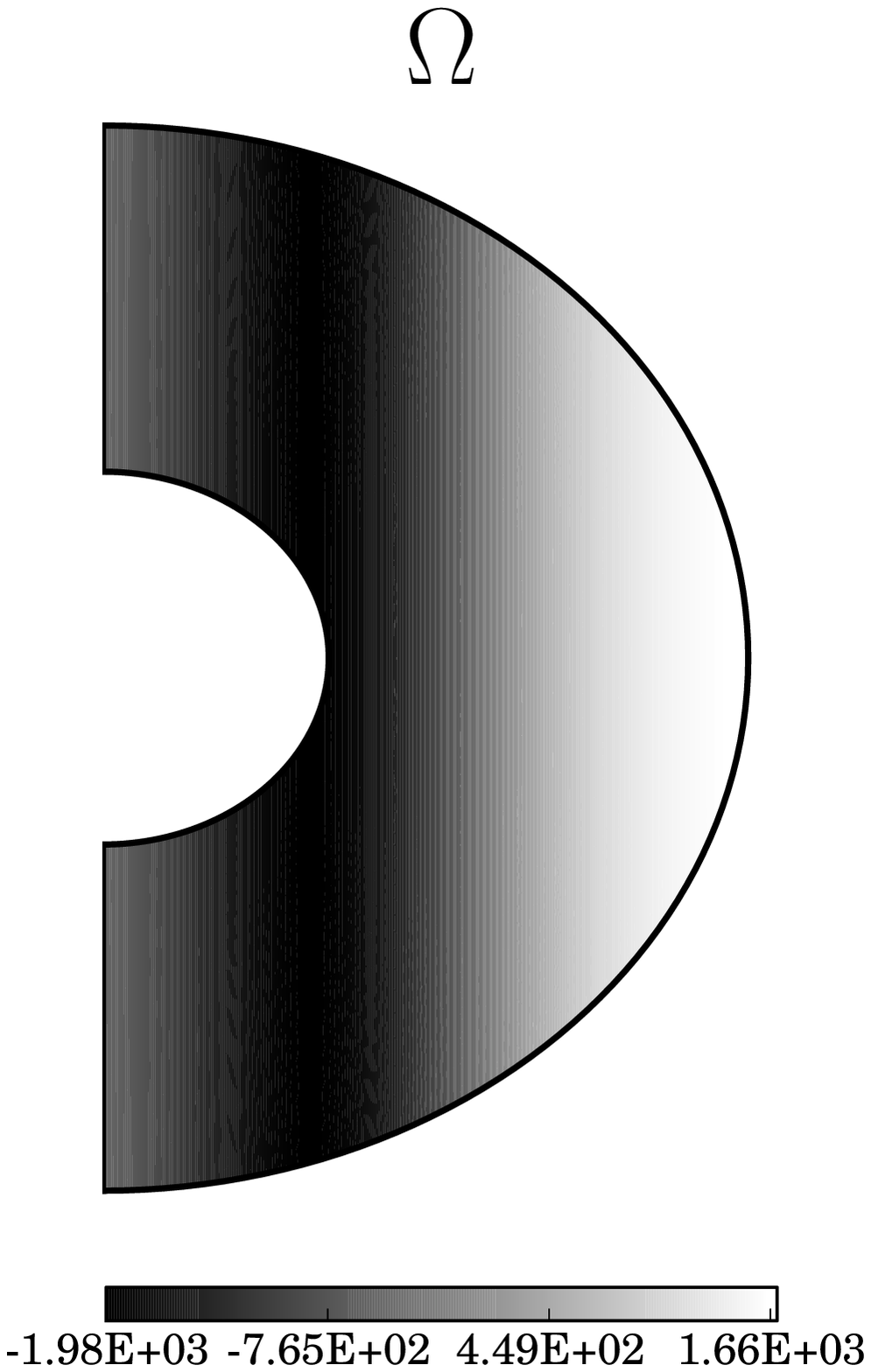}
\vskip -2cm
    \caption{The meridional sections of  $C_\alpha\alpha_\circ$, and  $C_\omega\Omega$ for  $C_\alpha=4.5 \, 10^2$ and $C_\omega=1.7\,10^3$.
} \label{fig_1}
\end{figure}

Approximations of the averaged on time axi-symmetrical angular velocity $\Omega$ and $\alpha=-\chi$ have the form:
\begin{equation}\label {Parker:5}
\begin{array}{l}
    \displaystyle
     \alpha=2.94 \, r (1-erf(1.25|z|))\,e^{-200/3\, (s - 0.39)^2}\sin(2\theta)\\ \\
    \Omega=-1.37\,e^{-11.77\, (s - 0.35)^2} + e^{-3.8\, (s - 1)^2}
\end{array}    
\end{equation}
with $s=r\sin\theta$, $z=r\cos\theta$, $V_\varphi=s\,\Omega$, see Fig.~\ref{fig_1}.

We stress attention that the  particular details of $\bf V$ and $\chi$ are sensitive to the parameters \cite{Simit:2004}, and other possible distributions of $\bf V$ and $\chi$ can be discussed.

 To analyze dependence of magnetic field properties on $C_\alpha$, and $C_\omega$,  we consider three mean  quantities: the 
  magnitude of the magnetic field dipole $\overline{g_1^0}$ ($g_l^0$ are  coefficients in the spectrum $\cal S$ on  the  Legendre polynomials $P_l$);
the  measure of $g_1^0$ oscillations, introduced as
   ${\cal R}=ln \left[\left(\overline{(g_1^0- \overline{g_1^0})^2}\right)^{1/2}/|\overline{g_1^0}|\right]$; and the number of the maximal harmonic $l_{\rm max}$ in the spectrum $\cal S$. Here ${\cal R}>0$ corresponds to the regime with the reversals of the field, when the magnetic dipole ($l=1$)  changes its sign. For  ${\cal R}<0$ reversals are absent, it  is the so-called regime in oscillations. 

  Each run for the certain pair of ($C_\alpha,\,C_\omega$) started from the initial seed of the magnetic field. In all the regimes magnetic 
   field come to the non-linear state after  the time less   $t=1$. After that 
 the mentioned above three quantities were averaged over the time interval $t=1.5-2$. 

\begin{figure}[th!] 
\def \ss {12cm}
\vskip -5.0cm
\hskip 0cm \includegraphics[width=\ss]{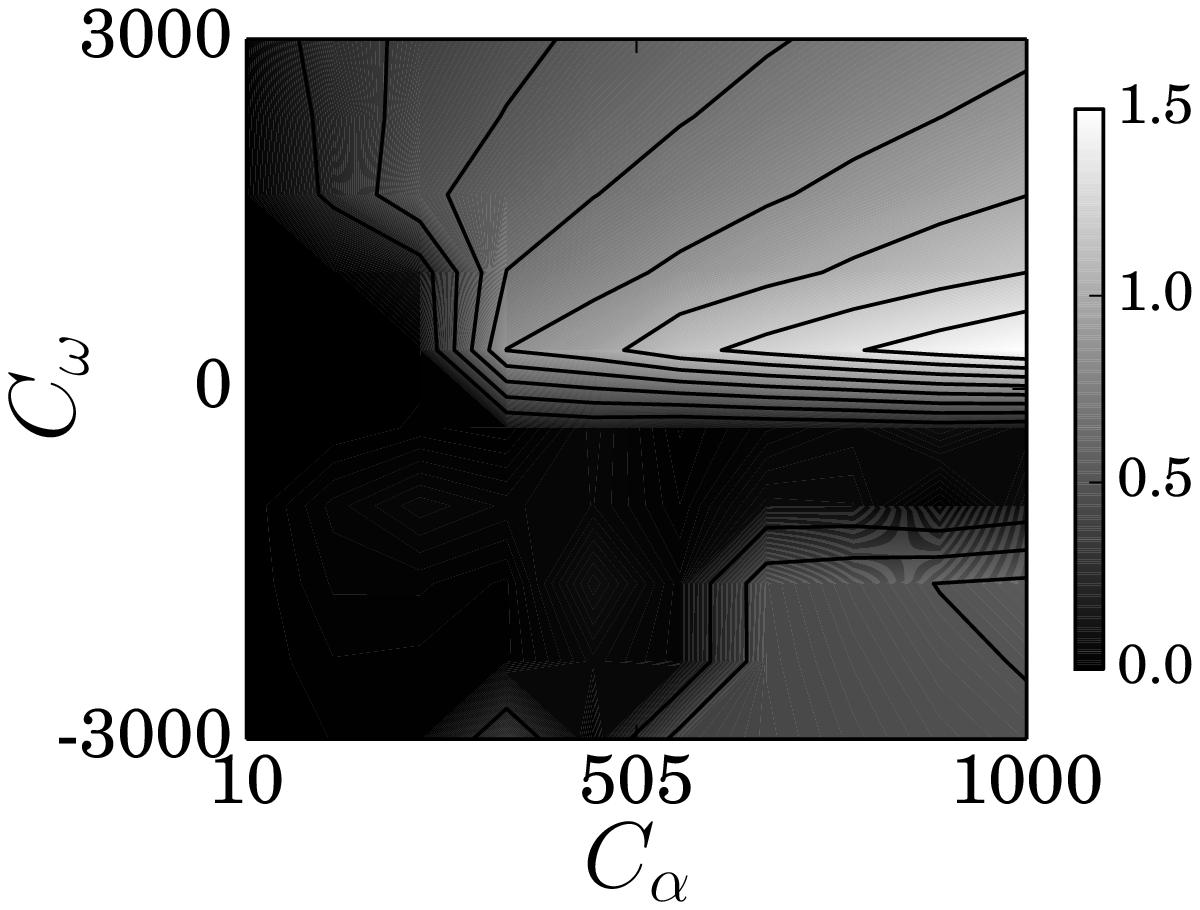}
\vskip -10.0cm
\hskip 0cm \includegraphics[width=\ss]{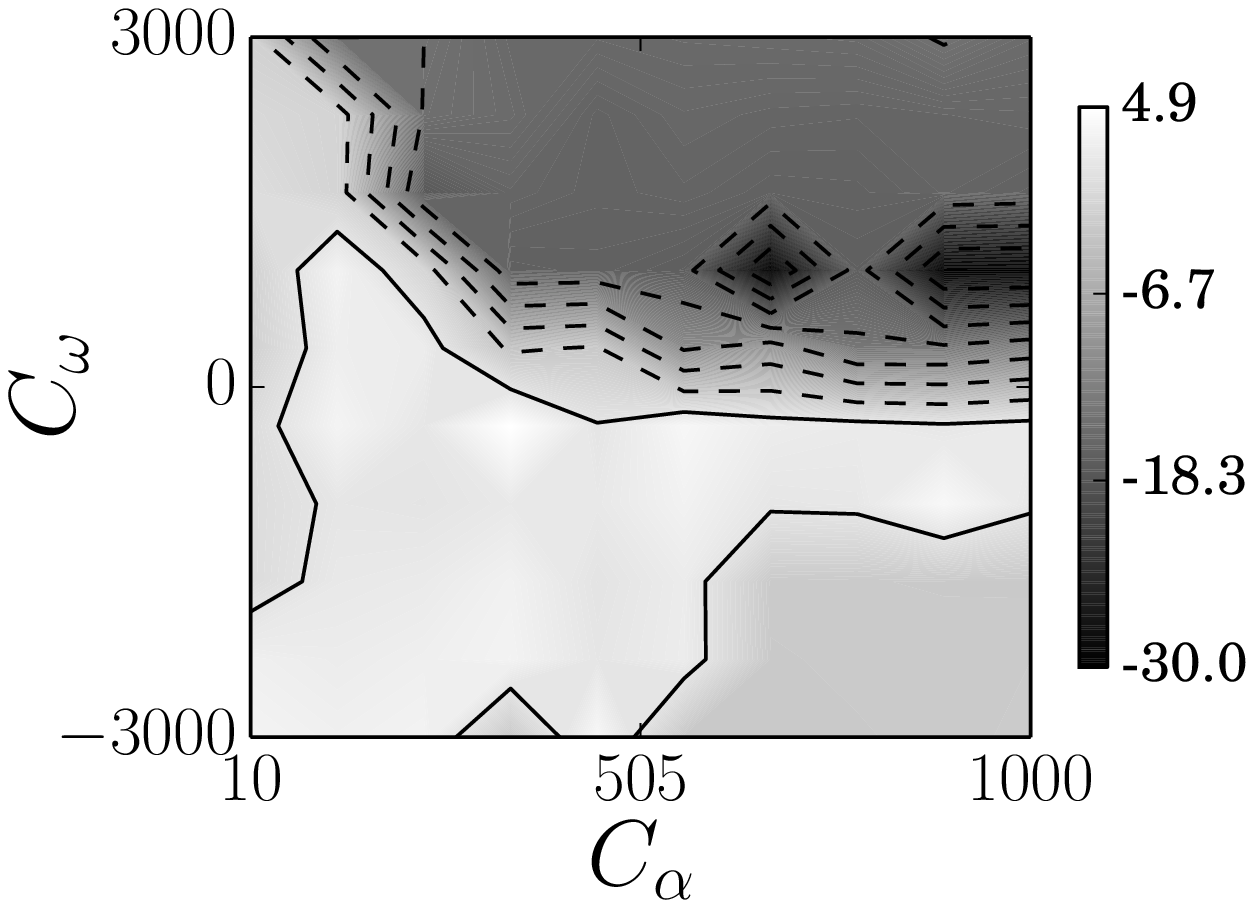}
\vskip -10.0cm
\hskip 0cm \includegraphics[width=\ss]{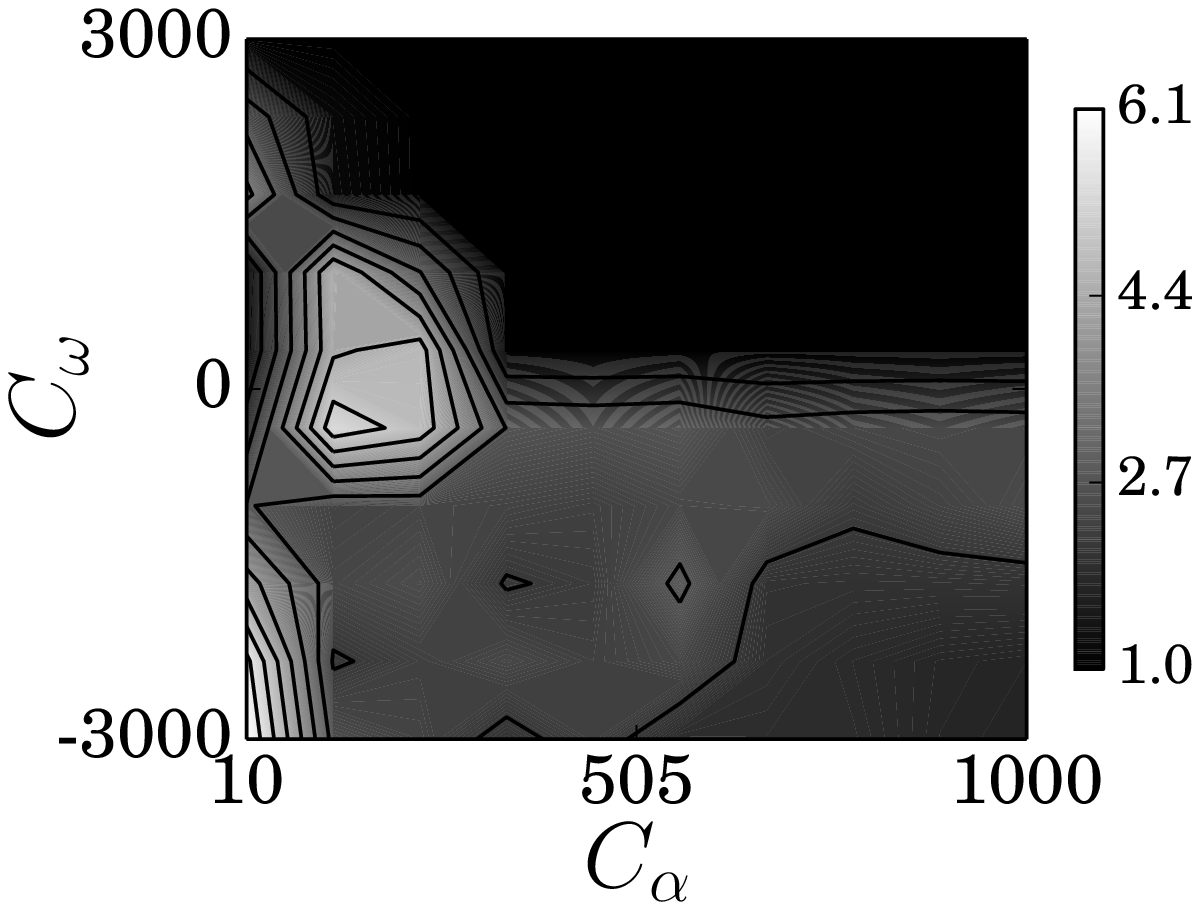}
\vskip -21.5 cm 
\hskip 9.5cm (a) 
\vskip 5cm 
\hskip 9.5cm (b) 
\vskip 5cm 
 \hskip 9.5cm (c) 
\vskip 5cm
  \caption{The dipole's $g_1^0$ amplitude (a), amplitude of the dipole variation $\cal R$ (b), and 
 the maximal number of harmonic $l_{\rm max}$ in the spectrum $S$ (c). All the quantities are averaged over the time.
} \label{fig_2}
\end{figure}

  As follows from analytics, input of the energy in system (\ref{Parker:3})  with $\eta=1$ is controlled by the product 
   ${\cal D}=C_\alpha\cdot C_\omega$, the dynamo-number. It corresponds to the hyperbolic structure of isolines in Fig.~\ref{fig_2}(a), however  some deviations from this prediction  still exist. Accordingly to 3D simulations \cite{Reshetnyak:2010}  $C_\alpha$ and $C_\omega$ are positive in geodynamo. As regards to the sign of the $\alpha$-effect, it  immediately follows from the negative sign of the kinetic  helicity $\chi$ in the northern hemisphere. The behavior of the azimuthal velocity $\Omega$ is more complex, because it depends on parameters, say the Rayleigh number, in more extent. 
     
 The system is not symmetric in respect to the change of the sign $\cal D$: generation at the positive $\cal D$ starts at the smaller values of $|{\cal D}|$.
    Also, as follows from the linear analysis (without non-linearity  (\ref{Parker:4})), change of the sign $\cal D$ leads to the change of oscillating mode to the stationary one and the dipole symmetry to the quadrupole symmetry, and vise versa. We observe different levels of oscillations in  Fig.~\ref{fig_2}(b), as well.

      Planetary magnetic fields are dipole. It means that even at the surface of the liquid core the first dipole  mode $g_1^0$ is larger than  the other harmonics with $l>1$. Due to the strong decay of the non-dipole modes in the thick mantle above the core  predominance of the dipole  at the surface of the Earth is only enforced.
     The number of  harmonic $l_{\rm max}$, which corresponds to the maximum in the spectrum at $r=1$, is plotted in Fig.~\ref{fig_2}(c). For positive $\cal D$, we do observe quite large region with the dipole dominating magnetic field. 
 
\begin{figure}[th!] 
\def \ss {7cm}
\vskip -1cm
 \includegraphics[width=\ss]{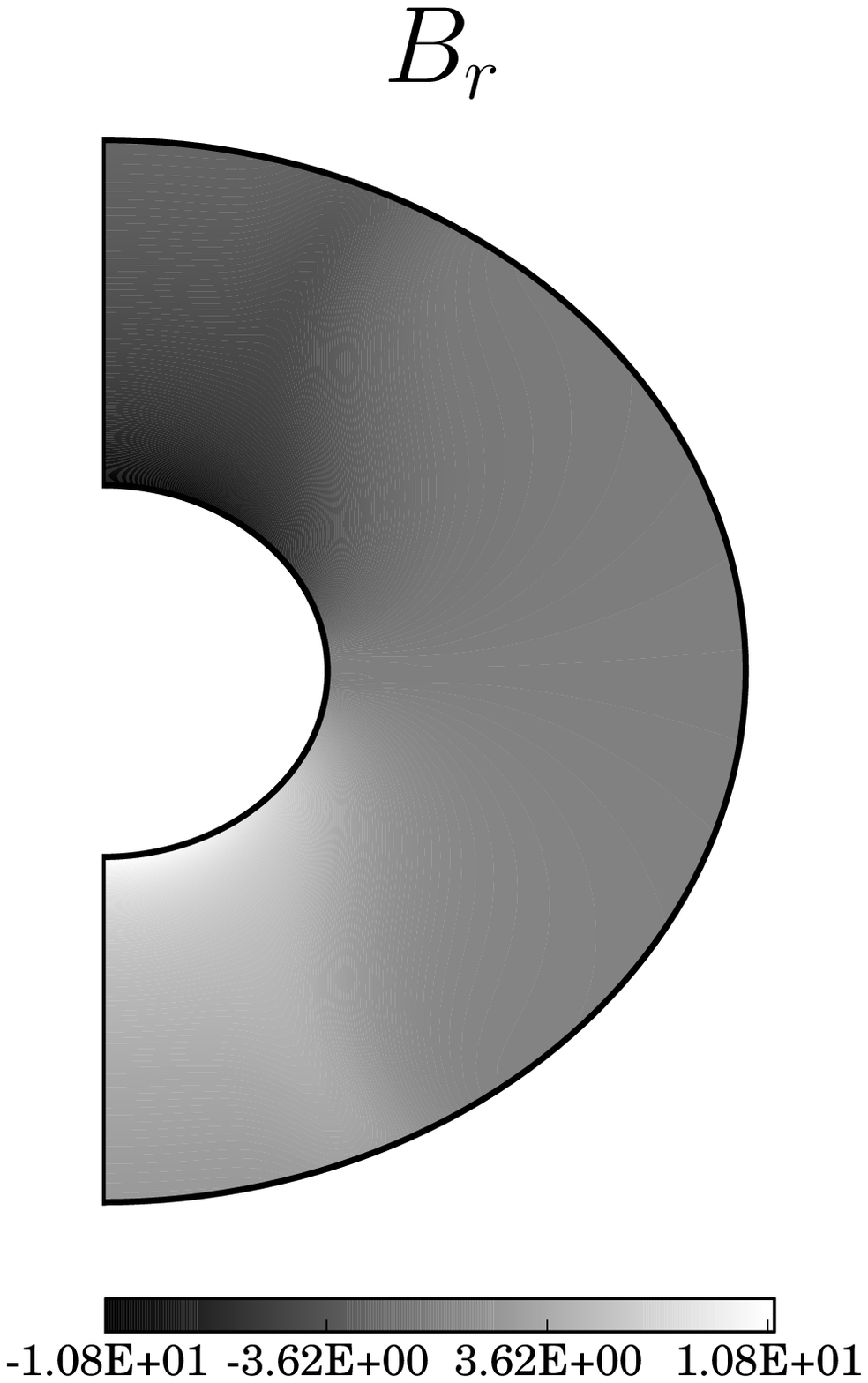}
\hskip -2cm \includegraphics[width=\ss]{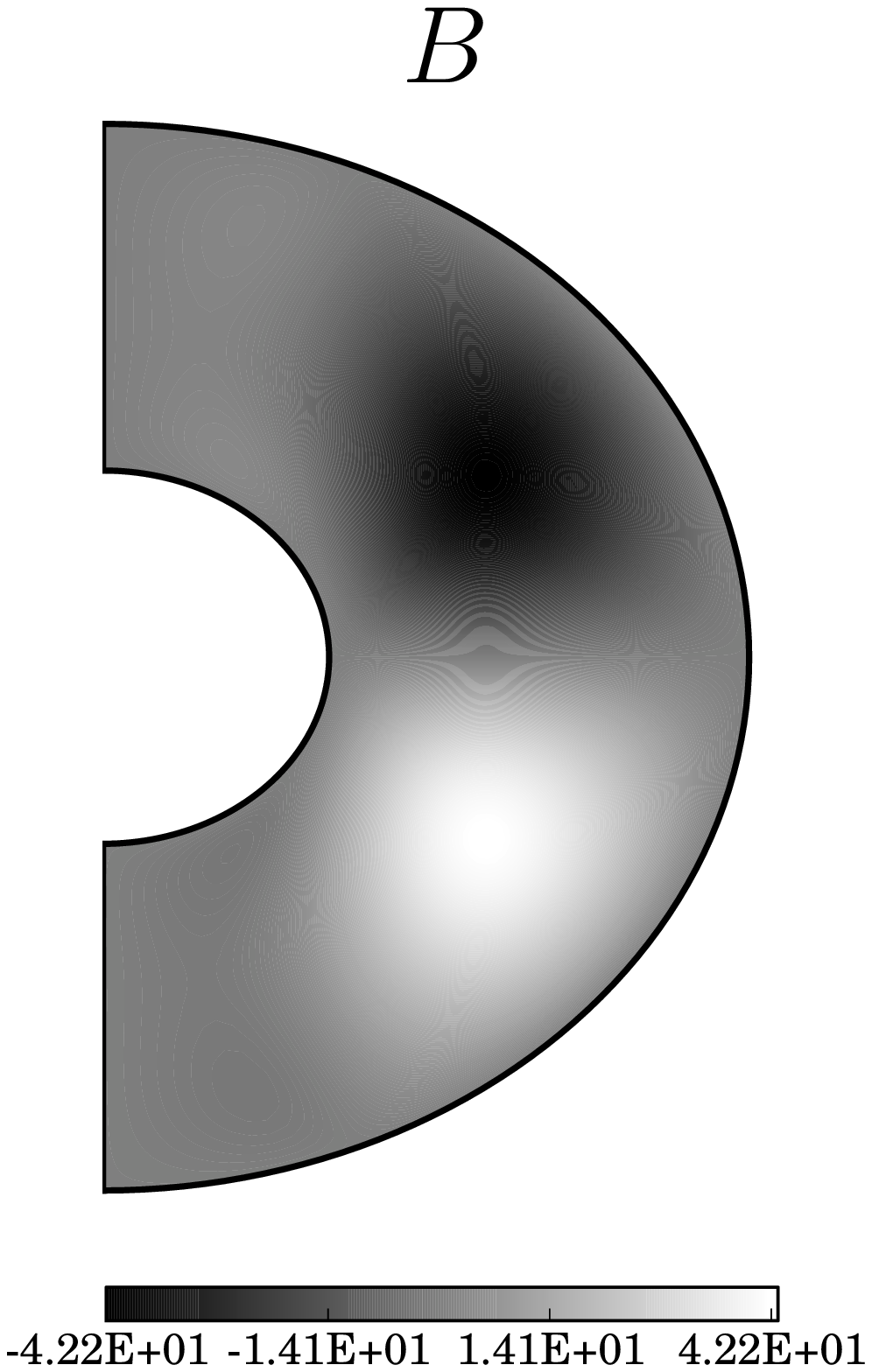}
\vskip -2cm
 \caption{Meridional sections of $B_r$- and $B$-components of the magnetic field.
} \label{fig_3}
\end{figure}

   The structure of the poloidal magnetic  field,  presented for the stationary regime  for  $C_\alpha=4.5 \, 10^2$, and $C_\omega=1.7\,10^3$, 
  see Fig.~\ref{fig_3},  
   resembles $Z$-profiles in Braginsky's model. However we get this result in a different manner, using profiles of $\alpha_\circ$ and $\Omega$,  taken from 3D simulations with the  
     simplest model of $\alpha$-quenching   (\ref{Parker:4}). Recall that in $Z$-model  additional two equations for the mean velocity field were solved. 

        In some sense existence of the magnetic field, trapped in the Taylor cylinder (TC), is predictable, because of the  geostriohic profiles of $\alpha_\circ$ and $\Omega$. This phenomenon is known in the full 3D simulations, where geostrophic convection in the spherical shell generates dipole dominating magnetic field. Moreover,   
   transition from the non-reversing magnetic field regime to the regime with reversals is controlled by the Rossby number $\rm Ro$ 
         \cite{Christensen:2006}. Accordingly to this scenario, increase of $\rm Ro$ leads to transition from the non-reversing dipole magnetic field regime to the regime with the reversals. In its turn, increase of $\rm Ro$ corresponds to attenuation of the cylindrical symmetry, concerned with the Coriolis force, and transition of the system towards the spherically symmetrical state, caused by the radial gravitational forces. In the latter case there is no any predominant direction, and magnetic field reverses frequently. Due to the lack of the mean kinetic helicity in the slow rotating systems, the large scale magnetic field, including the dipole one, is weak. What is interesting that maximum of the poloidal magnetic field $B_r$, which is inside of  TC,  is shifted 
      from the maximum of $\alpha_\circ$, located outside of   TC,  towards the axis.
   
\section{Random $\alpha$}  
   One of the difficulties of the mean-field dynamo is production of  stochastic sequences of the reversals, comparable to that ones in the paleomagnetic records.
          However in principle possibility of such regimes exists \cite{Hollerbach:1992}, in general, the mean-field dynamo, in contrast to the full  3D dynamo, tends to the stationary, either to quasi-periodic in time states. In the same time, it is known that the geomagnetic field  during the last 170My (and more) is characterized with the wide range of time interval lengths between the reversals: from $10^5$y to $10^7$y. The distribution of reversals over this time interval is not possible to approximate by the Poisson law with the constant parameters, and the best choice is to accept the hypotheses on the fractal nature of the reversals \cite{Anufriev:1994}, which help to include the superchrones of the magnetic field into the consideration.

         There are different possibilities how this contradiction can be overcome. The most trivial one is to accept modulation of the geodynamo parameters by the processes in the mantle. The other way is to find some special set of parameters near the bifurcation point, were  transition from superchrones  to the regimes with  the frequent reversals takes place. However, it is quite difficult to justify the choice of the particular parameters, close to the bifurcation point, itself, and only now it is realized that such points can be attractors \cite{Stefani:2006}. This fact tells in favor of probability increase of these regimes.
 
         Here we consider elegant approach, based on the idea of the stochastic nature of the mean-field parameters in the dynamo  equations. Due to the finite number of convective cells,  averaging of $\alpha$ and $\omega$ over the number of cells leads to  fluctuations \cite{Hoyng:1993}. The same happens when we consider averaging over the time using limited time series. Fluctuations  help to trigger the dynamo mechanism, and to produce reversals in the mean-field models. This approach is attractive  because it does not require additional mechanisms (like convection in mantle), and is capable to  produce stochastic geomagnetic polarity time scale \cite{Sobko:2012}. However we want to demonstrate that it should be used very carefully, because it can change the magnetic spectrum substantially.
 
              It is important that energy, introduced by $\alpha$-fluctuations,  will not propagate over the spectrum to the large scales. In stead of this, it will sink at the diffusion scale. It means that one can expect accumulation of the energy, concerned with fluctuations, at the scales not larger than the scale of the  energy injection. It is interesting that the mean-field dynamo is an example of  the self-organizing system, where energy of the turbulence is transformed to the energy of the large-scale magnetic field due to the included $\alpha$-effect. The inverse cascade due to $\alpha$-effect is made by hand.           
                But the mean-field equations itself can not provide the inverse cascade of energy for fluctuations of $\alpha$ introduced at some intermediate scale. Then, fluctuation of $\alpha$  transform magnetic spectrum in such a way that magnetic dipole will be smaller than the higher harmonics. In contrast to  2D \cite{KM:1980}, either quasi-geosrtophic convection \cite{RH:2008}, the  mean-field equations demonstrates the direct cascade of the energy. As a result, perturbations at the small scales do not contribute to the large-scale harmonics, and magnetic dipole stops to be a leading mode. This contradicts to the  paleomagnetic field observations, which demonstrate predominance of the dipole mode in the past.

 To demonstrate these ideas we modify (\ref{Parker:4}) in the following form:
  \begin{equation}\label{Parker:5}\displaystyle
    \alpha=C_\alpha\,{\alpha_\circ(r,\,\theta)\,(1+C_\delta\epsilon)   \over 1+E_m},
\end{equation}
  where $\epsilon$ is a uniformly distributed random variable in the interval $[-1, \, 1]$, and $C_\delta$ is a constant. Here $\epsilon$ is computed in every grid point, and modified after each 100 time steps $\tau=10^{-6}$. 

\begin{figure}[th!] 
\def \ss {12cm}
\vskip -5.0cm
\hskip 0cm \includegraphics[width=\ss]{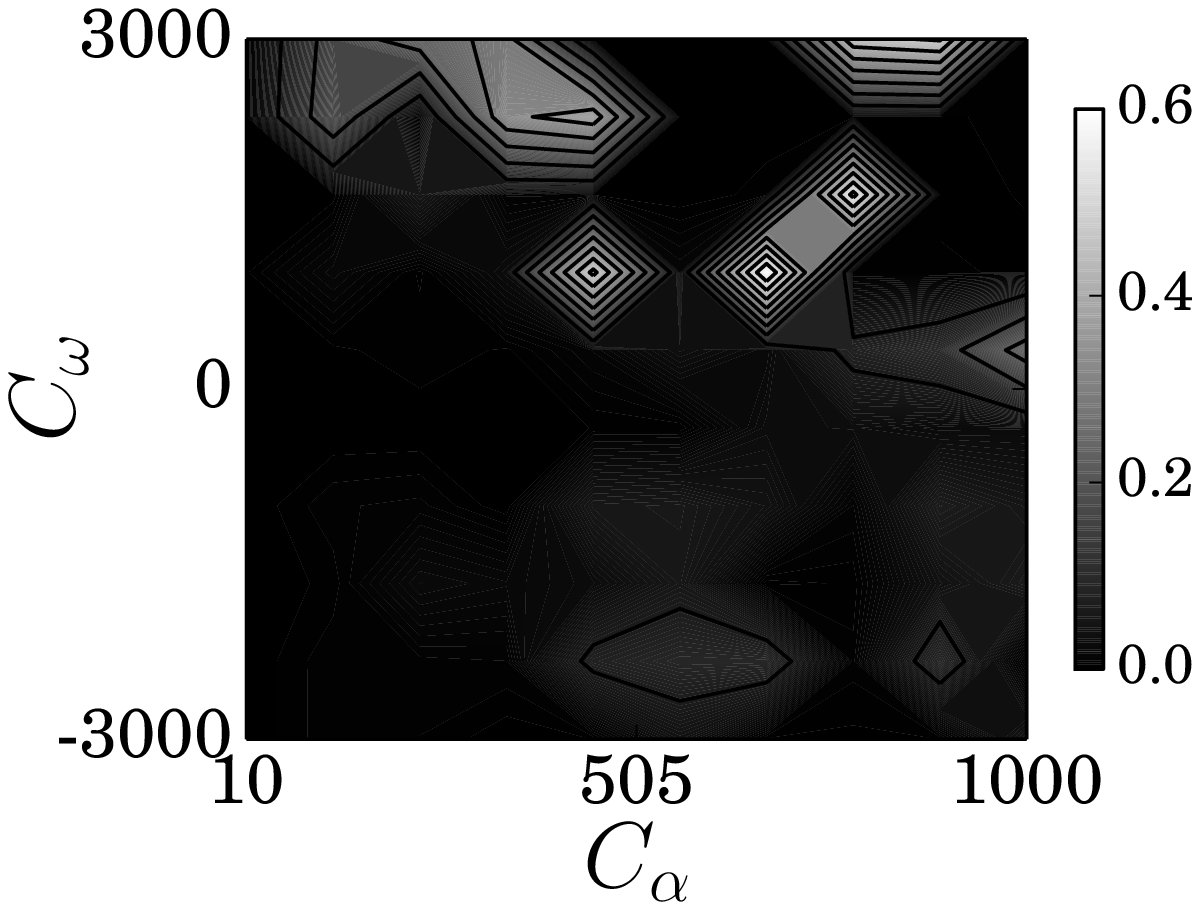}
\vskip -10.0cm
\hskip 0cm \includegraphics[width=\ss]{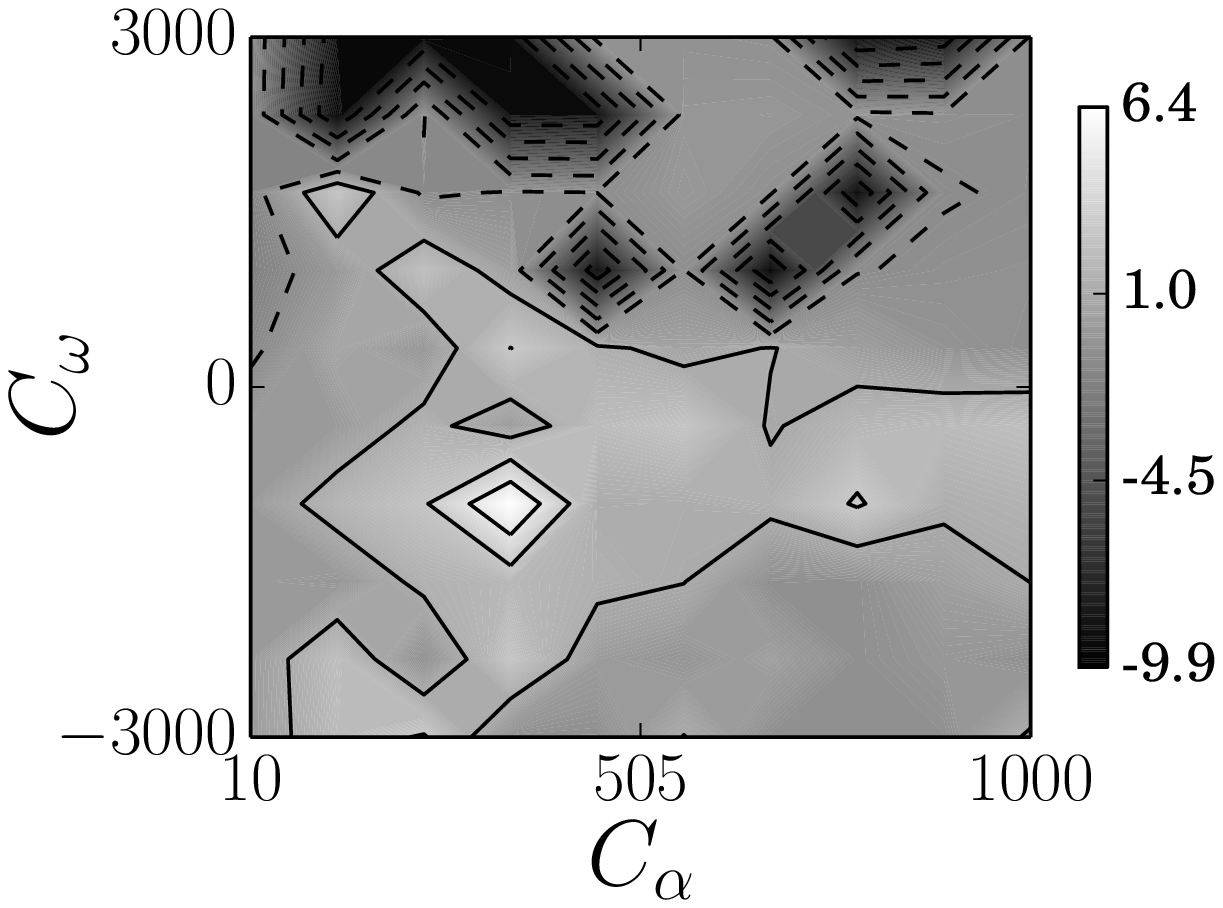}
\vskip -10.0cm
\hskip 0cm \includegraphics[width=\ss]{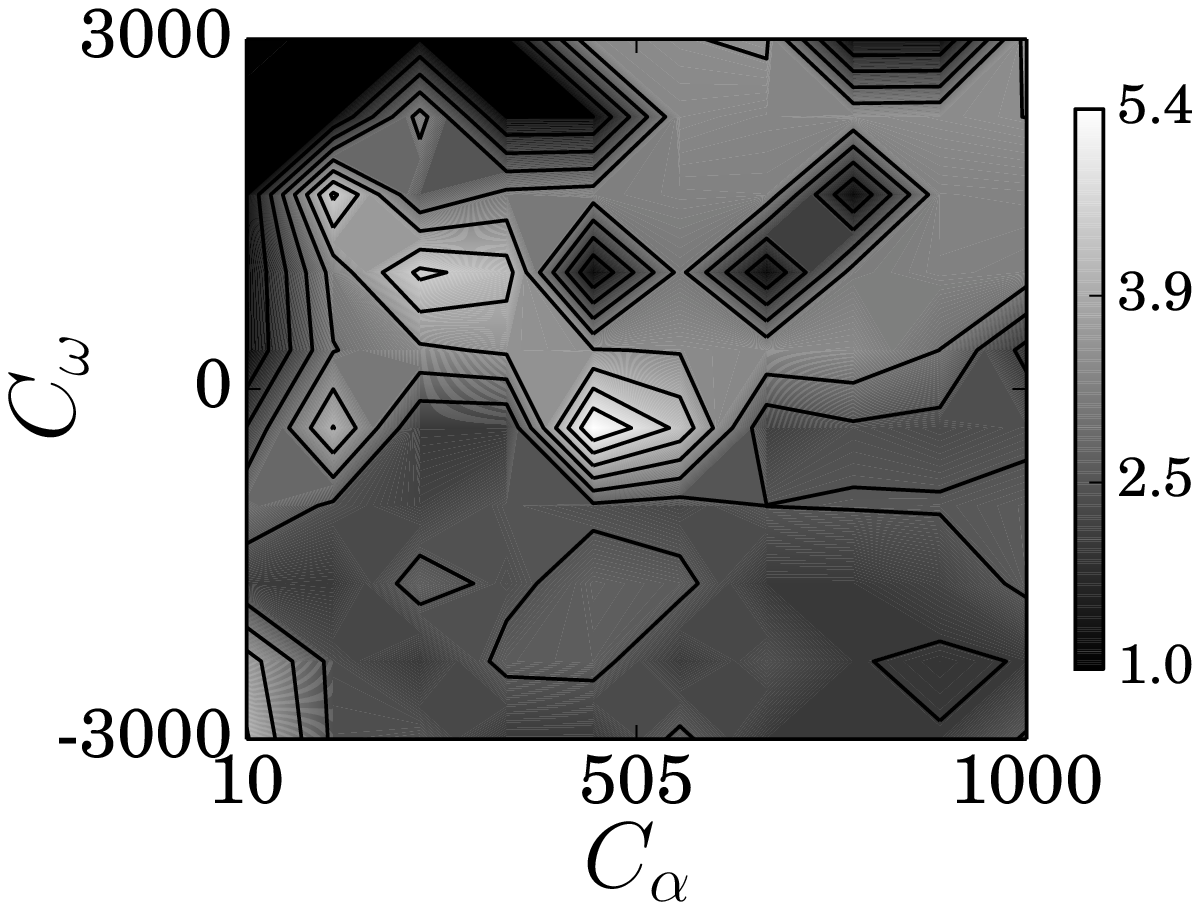}
\vskip -21.5 cm 
\hskip 9.5cm (a) 
\vskip 5cm 
\hskip 9.5cm (b) 
\vskip 5cm 
\hskip 9.5cm (c) 
\vskip 5cm
   \caption{Simulations with the noise, $C_\delta=1$. The dipole's $g_1^0$ amplitude (a), amplitude of the dipole variation $\cal R$ (b), and 
     the maximal number of harmonic $l_{\rm max}$ in the spectrum $S$ (c). 
} \label{fig_4}
\end{figure}

         To get substantial change of the reversal frequency, para\-metrized with $\cal R$,  we increased $C_\delta$, see Fig.~\ref{fig_4}. The first point is that fluctuations of $\alpha$ lead to decrease of the magnetic dipole amplitude, see Fig.~\ref{fig_4}(a). We do observe some islands with ${\cal R}>0$, where reversals take place, however they should be excluded for the  geodynamo applications because of the constraint  on the dipole domination in the magnetic spectrum. Usually all the regions with ${\cal R}>0$   correspond to $l_{\rm max}>1$, see Fig.~\ref{fig_4}(c).
  
  So far we do not know the exact values of 
  $(C_\alpha,\, C_\omega)$, our aim is to consider how fluctuations influence on the total area in maps  Fig.~\ref{fig_2}, \ref{fig_4}, which can
        be   associated with the frequent reversals regimes. From this point of view, fluctuations decrease probability of the dipole dominating magnetic field states, typical to the planetary dynamo. Anyway we can not exclude that such states do exist.
 

\section{Conclusions}   
    It is interesting that using a different from   $Z$-model's  profiles  of $\alpha$- and $\omega$-effects, and the form of quenching, the simulated  poloidal  magnetic field has the same  
    $Z$-structure. Sure, that in our model the inner core, which was absent in the original version of $Z$-model, plays important role, because 
    the $\alpha$-effect locates near the Stewartson layers, while in $Z$-model it was located in the narrow equatorial band. Note that taking into account compressibility of the liquid core, which produces kinetic helicity proportional to the product of the  radial gradient of density on the angular velocity of the planet,  will smooth the $\alpha$-profile over the whole volume even more. On contrary,  
          angular velocity in our model, and the geostrophic part of the angular velocity, which dominates over the other parts in $Z$-model, are very similar:
      near the axis rotation it is westward, and in the equator band it is eastward. 

                   As regards to  fluctuations of $\alpha$, we  intentionally considered the hard case, when $\alpha$ and $\omega$ are taken from the strong geostrophic state, where reversals are  improbable, so the amplitude of the noise $C_\delta$ must  be large. Increase of the Rossby number, which will attenuate geostrophy, can decrease $C_\delta$, and as a result, the change in the spectrum $\cal S$ will not be so sufficient. It also should be mentioned that  following strictly the basics of the mean-field dynamo, to derive the mean quantities one, should average 3D fields, using some intermediate scale, which is larger of the turbulent scale and smaller of the scale of the shell. This procedure, which was not done here,  smooths the geostrophic gradients, and helps to trigger the reversal's mechanism. Formally, it is equivalent to decrease of $\rm Ro$.

\Thanks{The authour acknowledges 
financial support from RFBR under grants 	15-05-00643, 15-52-53125.}


  \bibliographystyle{mhd}

\newcommand{\noopsort}[1]{}

\end{document}